 \documentclass[12pt]{article}

 \usepackage{amsmath, amssymb}
 \usepackage{eufrak, euscript}

\makeatletter

\renewcommand{\section}{\@startsection{section}{1}%
{\parindent}{3.6ex plus 0.8ex minus .2 ex} {1.6 ex plus .2 ex}{\large\bf}}   

\makeatother

\begin{document}

\title{On Weak Fields in Finsler Spaces}

\author{G.I. Garas'ko
\thanks{Russian Electrotechnical Institute, Moscow,
Russia, gri9z@mail.ru}}

\maketitle

\begin{abstract}
It is shown that in the weak field approximation the new geometrical
approach can lead to the linear field equations for the several independent
fields. For the stronger fields and in the second order approximation the
field equations become non-linear, and the fields become dependent. This
breaks the superposition principle for every separate field and produces the
interaction between different fields.The unification of the gravitational
and electromagnetic field theories is performed in frames of the geometrical
approach in the pseudo Riemannian space and in the curved Berwald-Moor space.
\end{abstract}

\section{Introduction}

In paper [1] the new (geometrical) approach was suggested for the field theory. It
is applicable for any Finsler space [2] for which in any point of the main space
${x^1,x^2,...,x^n}$ the indicatrix volume ${\left( \ \ V_{ind} (x^1,x^2,...,x^n)
\right) \ \ _{ev}}$ can be defined, provided the tangent space is Euclidean. Then
the action {$I$ }for the fields present in the metric function of the Finsler space
is defined within the accuracy of a constant factor as a volume of a certain
$n$-dimensional region {$V:$}
\begin{equation}
I=const\cdot {\int\limits_V}^{(n)}\frac{dx^1dx^2...dx^n}{\left(
V_{ind}(x^1,x^2,...,x^n)\right) _{ev}}\,.  \label{1}
\end{equation}
Thus, the field Lagrangean is defined in the following way
\begin{equation}
\EuScript{L}=const\cdot \frac 1{\left( V_{ind}(x^1,x^2,...,x^n)\right)
_{ev}}\,.  \label{2}
\end{equation}

In papers [3,4] the spaces conformally connected with the Minkowsky space
and with the Berwald-Moor space were regarded. These spaces have a single
scalar field for which the field equation was written and the particular
solutions were found for the spherical symmetry and for the
rhombododecaedron symmetry of the space.

The present paper is a continuation of those papers dealing with the study
and development of the geometric field theory.

\section{Pseudo Riemannian space with the signature (+,-,-,-)}

Let us consider the pseudo Riemannian space with the signature (+,-,-,-) and select
the Minkowsky metric tensor $\stackrel{o}{g}_{ij}$in the metric tensor,
${g_{ij}(x),}$ of this space explicitly
\begin{equation}
g_{ij}(x)=\stackrel{o}{g}_{ij}+h_{ij}(x)\,.  \label{3}
\end{equation}

Let us suppose that the field ${h_{ij}(x)}$ is weak, that is
\begin{equation}
|h_{ij}(x)|\ll 1\,.  \label{4}
\end{equation}

According to [1], the Lagrangean for a pseudo Riemannian space with the signature
(+,-,-,-) is equal to
 \begin{equation}
 \EuScript{L}=\sqrt{-det(g_{ij})}\,.  \label{5}
 \end{equation}

Let us calculate the value of ${  \left[ \ \ -det(g_{ij})\right] \ \ }$ within the
accuracy of ${  |h_{ij}(x)|^2:}$
 \begin{equation}
 -det(g_{ij})\simeq 1+\,\EuScript{L}_1+\,\EuScript{L}_2\,,  \label{6}
 \end{equation}
where
\begin{equation}
\EuScript{L}_1=\,\stackrel{o}{g}^{\,ij}h_{ij}\equiv
h_{00}-h_{11}-h_{22}-h_{33}\,,  \label{7}
\end{equation}
\begin{equation}
\vphantom{\frac{1}{2}}\EuScript{L}%
_2=-h_{00}(h_{11}+h_{22}+h_{33})+h_{11}h_{22}+h_{11}h_{33}+h_{22}h_{33}-h_{12}^2-h_{13}^2-h_{23}^2+h_{03}^2+h_{02}^2+h_{01}^2\,.
\label{8}
\end{equation}

The last formula can be rewritten in a more convenient way
 \begin{equation}
\begin{array}{l}
\EuScript{L}_2=-\left|
\begin{array}{cc}
h_{00} & h_{01} \\
h_{01} & h_{11}
\end{array}
\right| -\left|
\begin{array}{cc}
h_{00} & h_{02} \\
h_{02} & h_{22}
\end{array}
\right| -\left|
\begin{array}{cc}
h_{00} & h_{03} \\
h_{03} & h_{33}
\end{array}
\right| + \\
\\
\qquad \;\,+\left|
\begin{array}{cc}
h_{11} & h_{12} \\
h_{12} & h_{22}
\end{array}
\right| +\left|
\begin{array}{cc}
h_{11} & h_{13} \\
h_{13} & h_{33}
\end{array}
\right| +\left|
\begin{array}{cc}
h_{22} & h_{23} \\
h_{23} & h_{33}
\end{array}
\right| \,.
\end{array}
\label{9}
 \end{equation}
Then
 \begin{equation}
\EuScript{L}\simeq 1+\frac 12\,\EuScript{L}_1+\frac 12\,\left[ \EuScript{L}%
_2-\frac 14\,\EuScript{L}_1^2\right] \,,  \label{10}
 \end{equation}

To obtain the field equations in the first order approximation, one should use the
Lagrangean ${  \EuScript{L}_1,}$ and to do the same in the second
order approximation -- the Lagrangean ${  \left( \ \ \EuScript{L}_1+%
\EuScript{L}_2-\frac 14\,\EuScript{L}_1^2\right) \ \ .}$

\section{Scalar field}

For the single scalar field   $\varphi (x)$ the simplest representation of tensor {
$h_{ij}(x)$ }has the form{  \ }
 \begin{equation}
h_{ij}(x)\equiv h_{ij}^{(\varphi )}(x)=\pm \frac{\partial \varphi }{\partial
x^i}\frac{\partial \varphi }{\partial x^j}\,,  \label{11}
 \end{equation}

That is why  \
\begin{equation}
\EuScript{L}_\varphi =\sqrt{-det(g_{ij})}=\sqrt{1\pm \,\EuScript{L}_1}\simeq
1\pm \frac 12\,\EuScript{L}_1-\frac 18\,\EuScript{L}_1^2\,,  \label{12}
\end{equation}
where  \
\begin{equation}
\EuScript{L}_1=\left( \frac{\partial \varphi }{\partial x^0}\right)
^2-\left( \frac{\partial \varphi }{\partial x^1}\right) ^2-\left( \frac{%
\partial \varphi }{\partial x^2}\right) ^2-\left( \frac{\partial \varphi }{%
\partial x^3}\right) ^2\,.  \label{13}
\end{equation}

In the first order approximation, we can use the Lagrangean {  $%
\EuScript{L}_1$ }to obtain the following field equation
 \begin{equation}
\frac{\partial ^2\varphi }{\partial x^0\partial x^0}-\frac{\partial
^2\varphi }{\partial x^1\partial x^1}-\frac{\partial ^2\varphi }{\partial
x^2\partial x^2}-\frac{\partial ^2\varphi }{\partial x^3\partial x^3}=0\,,
\label{14}
 \end{equation}
which presents the wave equation. The stationary field that depends only on the
radius
\begin{equation}
r=\sqrt{(x^1)^2+(x^2)^2+(x^3)^2}\,,  \label{15}
\end{equation}
will satisfy the equation
 \begin{equation}
\frac d{dr}\left( r^2\frac{d\varphi }{dr}\right) =0\,,  \label{16}
 \end{equation}
the integration of which gives
\begin{equation}
\frac{d\varphi }{dr}=-\,C_1\frac 1{r^2}\qquad \Rightarrow \qquad \varphi
(r)=C_0+C_1\,\frac 1r\,.  \label{17}
\end{equation}

In the second order approximation one should use the Lagrangean {  $%
\left( \EuScript{L}_1-\frac 14\,\EuScript{L}_1^2\right) $ }to obtain the field
equation in the second order approximation
\begin{equation}
\stackrel{o}{g}^{\,ij}\frac \partial {\partial x^i}\left[ \left( \pm 1-\frac
12\,\EuScript{L}_1\right) \frac{\partial \varphi }{\partial x^j}\right] =0\,.
\label{18}
\end{equation}

This equation is already non-linear.

The strict field equation for the tensor ${  h_{ij}(x)}$ [11] is
\begin{equation}
\stackrel{o}{g}^{\,ij}\frac \partial {\partial x^i}\left( \frac{\displaystyle%
\frac{\partial \varphi }{\partial x^j}}{\sqrt{1\pm \,\EuScript{L}_1}}\right)
=0\,.  \label{19}
\end{equation}

Then the stationary field depending only on the radius must satisfy the equation
 \begin{equation}
\frac d{dr}\left( r^2\frac{\displaystyle\frac{d\varphi }{dr}}{\sqrt{1\mp
\,\left( \displaystyle\frac{d\varphi }{dr}\right) ^2}}\right) =0\,,
\label{20}
 \end{equation}
Integrating it, we get
\begin{equation}
\frac{d\varphi }{dr}=-\frac{C_1}{\sqrt{r^4\pm C_1^2}}\qquad \Rightarrow
\qquad \varphi (r)=C_0+\int\limits_r^\infty \frac{C_1}{\sqrt{r^4\pm C_1^2}}%
\,dr\,.  \label{21}
\end{equation}

The field with the upper sign and the field with the lower sign differ
qualitatively: the upper sign ''+'' in eq.(11) gives a finite field with no
singularity in the whole space, the lower sign ''-'' in eq.(11) gives a field
defined everywhere but for the spherical region
\begin{equation}
0\leq r\leq \sqrt{|C_1|}\,,  \label{22}
\end{equation}
in which there is no field, while
\begin{equation}
r>\sqrt{|C_1|}\,,\quad r\rightarrow \sqrt{|C_1|}\qquad \Rightarrow \qquad
\frac{d\varphi }{dr}\rightarrow -\,C_1\,\cdot \infty \,.  \label{23}
\end{equation}
At the same time in the infinity {  ($r\rightarrow \infty $) }both solutions {
$\varphi _{\pm }(r)$ }behave as the solution of the wave equation eq.(17).

If we know the Lagrangean, we can write the energy-momentum tensor, {  $%
T_j^i$, }for the obtained solutions and calculate the energy of the system devived
by the light speed,{  \ }$c$:
\begin{equation}
P_0=const\,\int^{(3)}T_0^0dV\,.  \label{24}
\end{equation}
To obtain the stationary spherically symmetric solutions, we get
\begin{equation}
T_0^0=-\,\frac{r^2}{\sqrt{r^4\pm C_1^2}}\,,  \label{25}
\end{equation}
that is why for both upper and lower signs   $P_0\rightarrow \infty $.

The metric tensor of eqs.(3,11) is the simplest way to ''insert'' the gravity field
into the Minkowsky space -- the initial flat space containing no fields. Adding
several such terms as in eq.(11) to the metric tensor, we can describe more and more
complicated fields by tensor $h_{ij}=h_{ij}^{(grav)}.$

\section{Covariant vector field}

To construct the twice covariant symmetric tensor ${  h_{ij}(x)}$ with the help of a
covariant field ${  A_i(x)}$ not using the connection objects, pay attention to the
fact that the alternated partial derivative of a tensor is a tensor too
 \begin{equation}
F_{ij}=\frac{\partial A_j}{\partial x^i}-\frac{\partial A_i}{\partial x^j}\,,
\label{26}
 \end{equation}
but a skew-symmetric one. Let us construct the symmetric tensor on the base of
tensor ${  F_{ij}.}$ To do this, first, form a scalar
\begin{equation}
\EuScript{L}_A=\stackrel{o}{g}^{\,ij}\stackrel{o}{g}^{\,km}F_{ik}F_{jm}=2%
\stackrel{o}{g}^{\,ij}\stackrel{o}{g}^{\,km}\left( \frac{\partial A_k}{%
\partial x^i}\frac{\partial A_m}{\partial x^j}-\frac{\partial A_k}{\partial
x^i}\frac{\partial A_j}{\partial x^m}\right) \,,  \label{27}
\end{equation}
which gives the following expressions for two symmetric tensors
\begin{equation}
h_{ij}^{(1)}=\stackrel{o}{g}^{\,km}\left( 2\frac{\partial A_k}{\partial x^i}%
\frac{\partial A_m}{\partial x^j}-\frac{\partial A_k}{\partial x^i}\frac{%
\partial A_j}{\partial x^m}-\frac{\partial A_k}{\partial x^j}\frac{\partial
A_i}{\partial x^m}\right) \,,  \label{28}
\end{equation}
\begin{equation}
h_{ij}^{(2)}=\stackrel{o}{g}^{\,km}\left( 2\frac{\partial A_i}{\partial x^k}%
\frac{\partial A_j}{\partial x^m}-\frac{\partial A_i}{\partial x^k}\frac{%
\partial A_m}{\partial x^j}-\frac{\partial A_j}{\partial x^k}\frac{\partial
A_m}{\partial x^i}\right) \,.  \label{29}
\end{equation}

Notice, that not only ${  F_{ij}\ }and{  \ \EuScript{L}_A,}$ but also the tensors ${
h_{ij}^{(1)},\ h_{ij}^{(2)}}$ are gradient invariant, that is they don't change with
transformations
\begin{equation}
A_i\rightarrow A_i+\frac{\partial f(x)}{\partial x^i}\,,  \label{30}
\end{equation}
where $f(x)${  \ }is an arbitrary scalar function{  . }

Let
\begin{equation}
h_{ij}\equiv h_{ij}^{(A_k)}=\chi (x)\,h_{ij}^{(1)}+\left[ 1-\chi (x)\right]
\,h_{ij}^{(2)}\vphantom{\frac{1}{2}}\,,  \label{31}
\end{equation}
where {  $\chi (x)$ }is some scalar function. Then in the first order approximation
we get
\begin{equation}
\EuScript{L}_1=2\stackrel{o}{g}^{\,ij}\stackrel{o}{g}^{\,km}\left( \frac{%
\partial A_k}{\partial x^i}\frac{\partial A_m}{\partial x^j}-\frac{\partial
A_k}{\partial x^i}\frac{\partial A_j}{\partial x^m}\right) \equiv %
\EuScript{L}_A\,,  \label{32}
\end{equation}
and the first order approximation for the field   $A_i(x)$ gives Maxwell equations
\begin{equation}
\stackrel{o}{g}^{\,ij}\frac{\partial ^2}{\partial x^i\partial x^j}A_k-\frac
\partial {\partial x^k}\left( \stackrel{o}{g}^{\,ij}\frac{\partial A_j}{%
\partial x^i}\right) =0\,.  \label{33}
\end{equation}

For Lorenz gauge
\begin{equation}
\stackrel{o}{g}^{\,ij}\frac{\partial A_j}{\partial x^i}=0\,,  \label{34}
\end{equation}
the equations eqs.(33) take the form
\begin{equation}
\square A_k=0\,.  \label{35}
\end{equation}

It is possible that eq.(31) is not the most general form for tensor {  $%
h_{ij}$ }which in the first order approximation gives the field equations
coinciding with Maxwell equations.

To obtain Maxwell equations not for the free field but for the field with sources
$j_i(x),$ one should add to ${  h_{ij}^{(A_k)}}$ the following tensor
\begin{equation}
h_{ij}^{(j_k)}=\left( \frac{16\pi }c\right) \cdot \frac 12\left(
A_ij_j+A_jj_i\right) \,,  \label{36}
\end{equation}

This means that the metric tensor eq.(3) with tensor
\begin{equation}
h_{ij}=h_{ij}^{(Max)}\equiv h_{ij}^{(A_k)}+h_{ij}^{(j_k)}  \label{37}
\end{equation}
describes the weak electromagnetic field with the source ${  \ j_k(x).}$ We must
bear in mind that we use the geometrical approach to the field theory, and we have
to consider ${  j_k(x)}$ to be given and not obtained from the field equations.

So, the metric tensor eq.(3) with tensor
\begin{equation}
h_{ij}=\mu h_{ij}^{(A_k)}+\gamma h_{ij}^{(grav)}\,,  \label{38}
\end{equation}
where $\mu ,\gamma $ are the fundamental constants in frames of the unique pseudo
Riemannian geometry describes simultaneously the free electromagnetic field and the
free gravitational field. To include the sources, $ j_k(x),$ of the electromagnetic
field, the metric tensor must either include not only {  $j_k(x)$ }but{  \ }the
partial derivatives of this field too or the field { $j_k(x)$ }must be expressed by
the other fields as shown below.

If the gravity field is ''inserted'' in the simplest way as shown in the previous
section, then the sources of the electromagnetic field can be expressed by the
scalar field as follows
\begin{equation}
j_i(x)=q\frac{\partial \varphi }{\partial x^i}\,.  \label{39}
\end{equation}
In this case the first order approximation for Lorenz gauge gives
\begin{equation}
\square A_k=\frac{4\pi }cj_k\,,  \label{40}
\end{equation}
\begin{equation}
\square \varphi =0\,.  \label{41}
\end{equation}
Since the density of the current has the form of eq.(39), the eq.(41) gives the
continuity equation
\begin{equation}
\stackrel{o}{g}^{\,ij}\frac{\partial j_i}{\partial x^j}=0\,.  \label{42}
\end{equation}

\section{Several weak fields}

The transition from the weak fields to the strong fields may lead to the transition
from the linear equations for the independent fields to the non-linear field
equations for the mutually dependent interacting fields {  $\varphi (x)$ }and{ \
$\psi (x)$ }''including'' gravity field in the Minkowsky space.

Let
\begin{equation}
h_{ij}=\varepsilon _\varphi \frac{\partial \varphi }{\partial x^i}\frac{%
\partial \varphi }{\partial x^j}+\varepsilon _\psi \frac{\partial \psi }{%
\partial x^i}\frac{\partial \psi }{\partial x^j},  \label{43}
\end{equation}
where {  $\varepsilon _\varphi $, $\varepsilon _\psi $ }are independent sign
coefficients. Then the strict Lagrangean can be written as follows
\begin{equation}
\EuScript{L}_{\varphi ,\psi }=\sqrt{1+\EuScript{L}_1+\EuScript{L}_2}\,,
\label{44}
\end{equation}
where  \
\begin{equation}
\EuScript{L}_1=\stackrel{o}{g}^{\,ij}\left( \varepsilon _\varphi \frac{%
\partial \varphi }{\partial x^i}\frac{\partial \varphi }{\partial x^j}%
+\varepsilon _\psi \frac{\partial \psi }{\partial x^i}\frac{\partial \psi }{%
\partial x^j}\right) \,,  \label{45}
\end{equation}
and  \
 \begin{equation}
\left.
\begin{array}{l}
\EuScript{L}_2=\varepsilon _\varphi \,\varepsilon _\psi \left[ -\displaystyle%
\left( \frac{\partial \varphi }{\partial x^0}\frac{\partial \psi }{\partial
x^1}-\frac{\partial \varphi }{\partial x^1}\frac{\partial \psi }{\partial x^0%
}\right) ^2+\right. \\
\\
\qquad \qquad -\displaystyle\left( \frac{\partial \varphi }{\partial x^0}%
\frac{\partial \psi }{\partial x^2}-\frac{\partial \varphi }{\partial x^2}%
\frac{\partial \psi }{\partial x^0}\right) ^2-\left( \frac{\partial \varphi
}{\partial x^0}\frac{\partial \psi }{\partial x^3}-\frac{\partial \varphi }{%
\partial x^3}\frac{\partial \psi }{\partial x^0}\right) ^2+ \\
\\
\qquad \qquad +\displaystyle\left( \frac{\partial \varphi }{\partial x^1}%
\frac{\partial \psi }{\partial x^2}-\frac{\partial \varphi }{\partial x^2}%
\frac{\partial \psi }{\partial x^1}\right) ^2+\left( \frac{\partial \varphi
}{\partial x^1}\frac{\partial \psi }{\partial x^3}-\frac{\partial \varphi }{%
\partial x^3}\frac{\partial \psi }{\partial x^1}\right) ^2+ \\
\\
\qquad \qquad \qquad \qquad \qquad \qquad \qquad \quad \;\,+\displaystyle%
\left. \left( \frac{\partial \varphi }{\partial x^2}\frac{\partial \psi }{%
\partial x^3}-\frac{\partial \varphi }{\partial x^3}\frac{\partial \psi }{%
\partial x^2}\right) ^2\right] \,.
\end{array}
\right\}  \label{46}
 \end{equation}

The expression eq.(46){  \ }can be obtained from eq.(9) most easily, if one uses the
following simplifying formula
\begin{equation}
\left|
\begin{array}{ll}
h_{ii_{-}} & h_{i_{-}j_{-}} \\
h_{i_{-}j_{-}} & h_{jj_{-}}
\end{array}
\right| =\pm \left|
\begin{array}{ll}
\displaystyle\frac{\partial \varphi }{\partial x^i} & \displaystyle\frac{%
\partial \psi \;\;}{\partial x^{i_{-}}} \\
&  \\
\displaystyle\frac{\partial \varphi \;\;}{\partial x^{j_{-}}} & \displaystyle%
\frac{\partial \psi }{\partial x^j}
\end{array}
\right| ^2=\pm \left( \frac{\partial \varphi }{\partial x^i}\frac{\partial
\psi }{\partial x^j}-\frac{\partial \varphi }{\partial x^j}\frac{\partial
\psi }{\partial x^i}\right) ^2\,.  \label{47}
\end{equation}

In the first order approximation for the Lagrangean, the expression $\EuScript{L}_1$
should be used. Then the field equations give the system of two independent wave
equations
 \begin{equation}
\left.
\begin{array}{c}
\displaystyle\frac{\partial ^2\varphi }{\partial x^0\partial x^0}-\frac{%
\partial ^2\varphi }{\partial x^1\partial x^1}-\frac{\partial ^2\varphi }{%
\partial x^2\partial x^2}-\frac{\partial ^2\varphi }{\partial x^3\partial x^3%
}=0\,, \\
\\
\displaystyle\frac{\partial ^2\psi }{\partial x^0\partial x^0}-\frac{%
\partial ^2\psi }{\partial x^1\partial x^1}-\frac{\partial ^2\psi }{\partial
x^2\partial x^2}-\frac{\partial ^2\psi }{\partial x^3\partial x^3}=0\,.
\end{array}
\right\}  \label{48}
 \end{equation}

Here the fields ${  \varphi (x)\ }and{  \ \psi (x)}$ are independent and the
superposition principle is fulfilled.

Using the strict Lagrangean for the two scalar fields eq.(44), we get a system of
two non-linear differential equations of the second order
\begin{equation}
\left.
\begin{array}{c}
\stackrel{o}{g}^{\,ij}\displaystyle\frac \partial {\partial x^i}\left[ \frac{%
\displaystyle\frac{\partial \varphi }{\partial x^j}\left( 1\pm \stackrel{o}{g%
}^{\,rs}\frac{\partial \psi }{\partial x^r}\frac{\partial \psi }{\partial x^s%
}\right) \mp \frac{\partial \psi }{\partial x^j}\stackrel{o}{g}^{\,rs}\frac{%
\partial \varphi }{\partial x^r}\frac{\partial \psi }{\partial x^s}}{\sqrt{1+%
\EuScript{L}_1+\EuScript{L}_2}}\right] =0\,, \\
\\
\stackrel{o}{g}^{\,ij}\displaystyle\frac \partial {\partial x^i}\left[ \frac{%
\displaystyle\frac{\partial \psi }{\partial x^j}\left( 1+\stackrel{o}{g}%
^{\,rs}\frac{\partial \varphi }{\partial x^r}\frac{\partial \varphi }{%
\partial x^s}\right) -\frac{\partial \varphi }{\partial x^j}\stackrel{o}{g}%
^{\,rs}\frac{\partial \varphi }{\partial x^r}\frac{\partial \psi }{\partial
x^s}}{\sqrt{1+\EuScript{L}_1+\EuScript{L}_2}}\right] =0\,.
\end{array}
\right\}  \label{49}
\end{equation}

Here the fields {  $\varphi (x)$, $\psi (x)$ }depend on each other, and the
superposition principle is not fulfilled. The transition from eqs.(48) to eqs.(49)
may be regarded as the transition from the weak fields to the strong fields.

\section{Non-degenerate polynumbers}

Consider a certain system of the non-degenerate polynumbers ${  P_n}$ [5], that is
$n$-dimensional associative commutative non-degenerated hyper complex numbers. The
corresponding coordinate space ${  x^1,x^2,...,x^n}$ is a Finsler metric flat space
with the length element equal to
 \begin{equation}
 ds=\displaystyle\sqrt[n]{\stackrel{o}{g}
_{i_1i_2...i_n}dx^{i_1}dx^{i_2}...dx^{i_n}}\,,  \label{50}
 \end{equation}
$\stackrel{o}{g}_{i_1i_2...i_n}$ is the metric tensor which does not depend on the
point of the space. The Finsler spaces of this kind can be found in literature (e.g.
[6,9]) but the fact that all the non-degenerated polynumber spaces belong to this
type of Finsler spaces was established beginning from the papers [10,11] and the
subsequent papers of the same authors, especially in [5].

The components of the generalized momentum in geometry corresponding to eq.(50) can
be found by the formulas
\begin{equation}
p_i=\displaystyle\frac{\stackrel{o}{g}_{ij_2...j_n}dx^{j_2}...dx^{j_n}}{%
\displaystyle\left( \stackrel{o}{g}%
_{i_1i_2...i_n}dx^{i_1}dx^{i_2}...dx^{i_n}\right) ^{\frac{n-1}n}}\,.
\label{51}
\end{equation}

The tangent equation of the indicatrix in the space of the non-degenerated
polynumbers ${  P_n}$ can be always written [5] as
 \begin{equation}
\stackrel{o}{g}^{\;i_1i_2...i_n}p_{i_1}p_{i_2}...p_{i_n}-\mu ^n=0\,,
\label{52}
 \end{equation}
where $\mu $ is a constatnt. There always can be found such a basis (and even
several such bases) and such a ${  \mu >0}$ that
 \begin{equation}
\left( \stackrel{o}{g}^{\,i_1i_2...i_n}\right) =\left( \stackrel{o}{g}%
_{\,i_1i_2...i_n}\right) \,.  \label{53}
 \end{equation}

Let us pass to a new Finsler geometry on the base of the space of non-degenerated
polynumbers ${  P_n}$. This new geometry is not flat, but its difference from the
initial geometry is infinitely small, and the length element in this new geometry is
 \begin{equation}
ds=\displaystyle\sqrt[n]{\left[ \stackrel{o}{g}_{i_1i_2...i_n}+\;\varepsilon
h_{i_1i_2...i_n}(x)\right] dx^{i_1}dx^{i_2}...dx^{i_n}}\,,  \label{54}
 \end{equation}
where $\varepsilon $ is an infinitely small value. If in the initial space the
volume element was defined by the formula
\begin{equation}
dV=dx^{i_1}dx^{i_2}...dx^{i_n}\,,  \label{55}
\end{equation}
in the new space within the accuray of $\varepsilon $ in the first power we have
 \begin{equation}
dV_h\simeq \left[ 1+\varepsilon \cdot C_0\stackrel{o}{g}^{%
\;i_1i_2...i_n}h_{i_1i_2...i_n}(x)\;\right] dx^{i_1}dx^{i_2}...dx^{i_n}\,,
\label{56}
 \end{equation}

That is according to [1], the Lagrangean of the weak field in the space with the
length element eq.(54) in the first order approximation is
 \begin{equation}
\EuScript{L}_1=\;\stackrel{o}{g}^{\;i_1i_2...i_n}h_{i_1i_2...i_n}(x)\,.
\label{57}
 \end{equation}
This expression generalizes formula eq.(7).

\section{Hyper complex space {\protect\large $H_4$}}

In the physical (''orthonormal'' [5]) basis in which every point of the space is
characterized by the four real coordinates  $x^0,x^1$, $ x^2,x^3$ the fourth power
of the length element  \ $ds_{H_4}$ is defined by the formula
\begin{equation}
\begin{array}{l}
\left( ds_{H_4}\right) ^4\equiv \;\stackrel{o}{g}_{ijkl}dx^0dx^1dx^2dx^3= \\
\\
\qquad \quad \;\,=(dx^0+dx^1+dx^2+dx^3)(dx^0+dx^1-dx^2-dx^3)\times \\
\\
\qquad \quad \;\,\times (dx^0-dx^1+dx^2-dx^3)(dx^0-dx^1-dx^2+dx^3)= \\
\\[6pt]
\qquad \quad \;\,=(dx^0)^4+(dx^1)^4+(dx^2)^4+(dx^3)^4+8dx^0dx^1dx^2dx^3- \\
\\
\qquad \qquad -2(dx^0)^2(dx^1)^2-2(dx^0)^2(dx^2)^2-2(dx^0)^2(dx^3)^2- \\
\\
\qquad \qquad -2(dx^1)^2(dx^2)^2-2(dx^1)^2(dx^3)^2-2(dx^2)^2(dx^3)^2\,.
\end{array}
\label{58}
\end{equation}

Let us compare the fourth power of the length element $ds_{H_4}$ in the space of
polynumbers   $H_4$ with the fourth power of the length element $ds_{Min}$ in the
Minkowsky space
\begin{equation}
\begin{array}{l}
\left( ds_{Min}\right) ^4=(dx^0)^4+(dx^1)^4+(dx^2)^4+(dx^3)^4- \\
\\
\qquad \qquad -2(dx^0)^2(dx^1)^2-2(dx^0)^2(dx^2)^2-2(dx^0)^2(dx^3)^2- \\
\\
\qquad \qquad +2(dx^1)^2(dx^2)^2+2(dx^1)^2(dx^3)^2+2(dx^2)^2(dx^3)^2\,.
\end{array}
\label{59}
\end{equation}
This means
\begin{equation}
\begin{array}{l}
\left( ds_{H_4}\right) ^4=\left( ds_{Min}\right) ^4+8dx^0dx^1dx^2dx^3- \\
\\
\qquad \quad \;\;-4(dx^1)^2(dx^2)^2-4(dx^1)^2(dx^3)^2-4(dx^2)^2(dx^3)^2\,,
\end{array}
\label{60}
\end{equation}
and in the covariant notation we have
\begin{equation}
\left( ds_{H_4}\right) ^4=\;\left( \stackrel{o}{g}_{ij}\stackrel{o}{g}%
_{kl}+\;\frac 13\stackrel{o}{g^{\prime }}_{ijkl}-\,\stackrel{o}{G}%
_{ijkl}\right) dx^idx^jdx^kdx^l\,,  \label{61}
\end{equation}
where
 \begin{equation}
\stackrel{o}{g^{\prime }}_{ijkl}=\left\{
\begin{array}{l}
1\,,\quad \hbox{if indeces}\,\,i,j,k,l\,\,,\text{ are all
different} \\
0\,,\quad \text{else}
\end{array}
\right.  \label{62}
 \end{equation}
 \begin{equation}
\stackrel{o}{G}_{ijkl}=\left\{
\begin{array}{l}
1\,,\quad \hbox{if }\,\,i,j,k,l\neq 0\,\,\hbox{~and~}\,\,i=j\neq k=l, \\
\qquad \qquad \qquad \qquad \quad \hbox{or~~}\,i=k\neq j=l, \\
\qquad \qquad \qquad \qquad \quad \hbox{or~~}\,i=l\neq j=k; \\
0\,,\quad \text{else}
\end{array}
\right.  \label{63}
 \end{equation}

The tangent equation of the indicatrix in the  $H_4$ space can be written in the
physical basis as in [5]
\begin{equation}
(p_0+p_1+p_2+p_3)(p_0+p_1-p_2-p_3)(p_0-p_1+p_2-p_3)(p_0-p_1-p_2+p_3)-1=0\,,
\label{64}
\end{equation}
where $p_i$ are the generalized momenta
 \begin{equation}
p_i=\frac{\partial \;ds_{H_4}}{\partial (dx^i)}\,.  \label{65}
 \end{equation}
Comparing formula eq.(64) with formula eq.(65), we have
\begin{equation}
\stackrel{o}{g}^{\;ijkl}p_ip_jp_kp_l-1=0\,.  \label{66}
\end{equation}
Here
\begin{equation}
\stackrel{o}{g}^{\;ijkl}=\;\stackrel{o}{g}^{\;ij}\stackrel{o}{g}%
^{\;kl}+\;\frac 13\stackrel{o}{g^{\prime }}^{\;ijkl}-\,\stackrel{o}{G}%
^{\;ijkl}\,,  \label{67}
\end{equation}
and
\begin{equation}
\left( \stackrel{o}{g}^{\;ijkl}\right) =\left( \stackrel{o}{g}_{ijkl}\right)
\,,\quad \left( \stackrel{o}{g^{\prime }}^{\;ijkl}\right) =\left( \stackrel{o%
}{g^{\prime }}_{\;ijkl}\right) \,,\quad \left( \stackrel{o}{G}%
^{\;ijkl}\right) =\left( \stackrel{o}{G}_{ijkl}\right) \,.  \label{68}
\end{equation}

To get the Lagrangean for the weak field in the first order approximation, we have
to get tensor ${  h_{ijkl}}$ in eq.(57). In the simplified version it could be
splitted into two additive parts: gravitational part and electromagnetic part. The
gravitational part can be constructed analogously to Sections 3 and 5 with regard to
the possibility to use the two-index number tensors, since now tensors
$\stackrel{o}{g}^{\;ijkl},\ h_{ijkl}\ $have four indices. The construction of the
electromagnetic part should be regarded in more detail.

Since we would like to preserve the gradient invariance of the Lagrangean and to get
Maxwell equations for the free field in the ${  H_4}$ space, let us write the
electromagnetic part of tensor ${  h_{ijkl}}$ in the following way
\begin{equation}
h_{ijkl}^{A_k}=\chi (x)\,h_{ijkl}^{(1)}+\left[ 1-\chi (x)\right]
\,h_{ijkl}^{(2)}\,,  \label{69}
\end{equation}
where tensors ${  h_{ijkl}^{(1)},\ h_{ijkl}^{(2)}}$ are the tensors present in the
round brackets in the r.h.s. of formulas eqs.(28,29). Then
\begin{equation}
\EuScript{L}_A=\quad \stackrel{o}{g}^{\,ijkl}h_{ijkl}^{A_k}\,\equiv 2%
\stackrel{o}{g}^{\,ij}\stackrel{o}{g}^{\,km}\left( \frac{\partial A_k}{%
\partial x^i}\frac{\partial A_m}{\partial x^j}-\frac{\partial A_k}{\partial
x^i}\frac{\partial A_j}{\partial x^m}\right) \,.  \label{70}
\end{equation}

To obtain Maxwell equations not for the free field but for the field with a source
${ j_i(x),}$ one should add to the tensor ${  \ h_{ijkl}^{(A_k)}}$ eq.(69) the
following tensor
\begin{equation}
h_{ijkl}^{(j_k)}=\left( \frac{16\pi }c\right) \cdot \frac 16\left( 2A_ij_j%
\stackrel{o}{g}_{kl}-A_i\stackrel{o}{g}_{jk}j_l-j_i\stackrel{o}{g}%
_{jk}A_l\,,\right) \,,  \label{71}
\end{equation}
symmetrized in all indices, that is tensor
\[
h_{ijkl}=h_{ijkl}^{Max}\equiv h_{ijkl}^{(A_k)}+h_{(ijkl)}^{(j_k)}
\]
describes the weak electromagnetic field with the sources ${  \ j_i(x)}$ where
\begin{equation}
j_i=\sum\limits_bq_{(a)}\frac{\partial \psi _{(b)}}{\partial x^i}\,,
\label{72}
\end{equation}
and $\psi _{(b)}$ are the scalar components of the gravitational field.

To obtain the unified theory for the gravitational and electromagnetic fields one
should take the linear combination of tensor $h_{ijkl}^{(Max)}$ corresponding to the
electromagnetic field in the first order approximation, and tensor
$h_{ijkl}^{(grav)}$ corresponding to the gravitational field in the first order
approximation
\begin{equation}
h_{ijkl}=\mu h_{ijkl}^{(Max)}+\gamma h_{ijkl}^{(grav)}\,,  \label{73}
\end{equation}
where $\mu ,\gamma $ are constants. Tensor ${  h_{ijkl}^{(grav)}}$ may be, for
example, constructed in the following way
\begin{equation}
h_{ijkl}^{grav}=\sum\limits_{a=1}^N\varepsilon _{(a)}\frac{\partial \varphi
_{(a)}}{\partial x^i}\frac{\partial \varphi _{(a)}}{\partial x^j}\frac{%
\partial \varphi _{(a)}}{\partial x^k}\frac{\partial \varphi _{(a)}}{%
\partial x^l}+\sum\limits_{b=1}^M\epsilon _{(b)}\frac{\partial \psi _{(b)}}{%
\partial x^{(i}}\frac{\partial \psi _{(b)}}{\partial x^j}\;\stackrel{o}{g}%
_{kl)},  \label{74}
\end{equation}
where ${  \varepsilon _{(a)},\,\epsilon _{(b)}}$ are the sign coefficients, and
$\varphi _{(a)},\,\psi _{(b)}$ are the scalar fields. The whole number of scalar
fields is equal to $(N+M)${  . }

\section{Conclusion}

In this paper it was shown that the geometrical approach [1] to the field
theory in which there usually appear the non-linear and non-splitting field
equations could give a system of independent linear equations for the weak
fields in the first order approximation. When the fields become stronger the
superposition principle (linearity) breaks, the equations become non-linear
and the fields start to interact with each other. We may think that these
changes of the equations that take place when we pass from the weak fields
to the strong fields are due to the two mechanisms: first is the qualitative
change of the field equations for the free fields in the first order
approximation; second is the appearance of the additional field sources,
that is the generation of the field by the other fields.

In frames of the geometrical approach to the field theory [1] the unification of the
electromagnetic and gravitational fields is performed both for the four-dimensional
pseudo Riemannian space with metric tensor ${  g_{ij}(x)}$ and for the
four-dimensional curved Berwald-Moor space with metric tensor ${  g_{ijkl}(x).}$

\end{document}